\documentclass[aps,longbibliography,superscriptaddress,10pt]{revtex4-1}

\usepackage{graphicx}
\usepackage{color}
\usepackage{subfigure}
\usepackage{amsmath}
\usepackage{amssymb} 
\usepackage{color}

\usepackage[margin=1.25in]{geometry}

\begin{document}

\title{Capillary filtering of particles during dip coating}

\author{Alban Sauret} \email{asauret@ucsb.edu}
\affiliation{Department of Mechanical Engineering, University of California, Santa Barbara, CA 93106, USA}

\author{Adrien Gans}
\affiliation{Surface du Verre et Interfaces, UMR 125, CNRS/Saint-Gobain, 93303 Aubervilliers, France}
\affiliation{FAST, CNRS, Univ. Paris-Sud, Univ. Paris-Saclay, 91405 Orsay, France}

\author{B\'en\'edicte Colnet}
\affiliation{Surface du Verre et Interfaces, UMR 125, CNRS/Saint-Gobain, 93303 Aubervilliers, France}

\author{Guillaume Saingier}
\affiliation{Surface du Verre et Interfaces, UMR 125, CNRS/Saint-Gobain, 93303 Aubervilliers, France}

\author{Martin Z. Bazant}
\affiliation{Department of Chemical Engineering, Massachusetts Institute of Technology, Cambridge, MA 02139, USA}
\affiliation{Department of Mathematics, Massachusetts Institute of Technology, Cambridge, MA 02139, USA}

\author{Emilie Dressaire} 
\affiliation{Department of Mechanical Engineering, University of California, Santa Barbara, CA 93106, USA}

\date{10 May 2019}

\begin{abstract}
An object withdrawn from a liquid bath is coated with a thin layer of liquid. Along with the liquid, impurities such as particles present in the bath can be transferred to the withdrawn substrate. Entrained particles locally modify the thickness of the film, hence altering the quality and properties of the coating. In this study, we show that it is possible to entrain the liquid alone and avoid contamination of the substrate, at sufficiently low withdrawal velocity in diluted suspensions. Using a model system consisting of a plate exiting a liquid bath, we observe that particles can remain trapped in the meniscus which exerts a resistive capillary force to the entrainment. We characterize different entrainment regimes as the withdrawal velocity increases: from a pure liquid film, to a liquid film containing clusters of particles and eventually individual particles. This capillary filtration is an effective barrier against the contamination of substrates withdrawn from a polluted bath and find application against bio-contamination.
\end{abstract}


\maketitle

\section{Introduction}

A solid body dipped in a liquid bath emerges covered with a thin layer of liquid. This familiar phenomenon occurs, for instance, when withdrawing a probe from a biological sample, but it is also essential in many industrial settings \cite{rio2017withdrawing}. Dip coating is commonly used to cover surfaces with a liquid layer of constant thickness, from several to hundreds of micrometers, depending on the conditions \cite{ruschak1985coating}. The thickness of the entrained film depends on the competition between the capillary pressure gradient and the viscous force at the liquid/solid interface through several factors such as the withdrawal speed, the physical properties of the fluid and surface \cite{landau1942physicochim,Deryagin,rio2017withdrawing,Quere:1999fr}. Recent advances in materials science combine modified surfaces and complex fluids to develop new coatings, with interesting physical and optical properties. 

\medskip

For a perfectly wetting Newtonian fluid, the first solution describing the thickness of the liquid film was proposed by Landau and Levich \cite{landau1942physicochim} and then Deryagin \cite{Deryagin}. Most of the research on liquid film entrainment has since focused on the great diversity of homogenous fluids and substrates and their influence on the film properties \cite{Mayer:2012dt}. Indeed, the solutions to the hydrodynamic problem and the film features depend on the nature of the fluid, through bulk properties and boundary conditions. As non-Newtonian fluids are commonly used in industrial applications, studies have also investigated the influence of the rheological properties on the film thickness \cite{maillard2014solid,maillard2015flow}. Surfactants modify the interfacial properties of the fluid and the result of the dip-coating process. Experiments show that surfactants increase the thickness of the entrained liquid layer by modifying the flow in the liquid bath \cite{park1991effects,shen2002fiber,krechetnikov2005experimental,krechetnikov2006surfactant,delacotte2012plate}.

\medskip

Much less emphasis has been placed on complex or heterogeneous fluids. In particular, a liquid bath can contain impurities in the form of rigid or soft particles, which may be entrained in the coating. For instance, microorganisms, sediments, and fragments of plastics are dispersed in natural water bodies and can easily contaminate solid surfaces. This transfer of particles is useful for biological processes and environmental measurements, but can also contribute to the dissemination of pathogens and pollutants. It is thus essential to be able to predict the conditions under which a particle dispersed in a liquid phase will be entrained in the liquid film and contaminate the substrate. Previous work on dip coating of suspensions has evidenced the entrainment of colloidal particles by a liquid film \cite{ghosh2007spontaneous,watanabe2009mechanism,brewer2011coating}. After withdrawal, the colloidal particles form complex patterns on the solid surface. The geometry of the patterns depends on the particle concentration, withdrawal speed, and evaporation of the liquid. Despite these recent advances, the transfer of dilute particles whose size is comparable to the film thickness continues to challenge the continuum description that applies to coated films of colloidal suspensions. 

\medskip

Over the past few years, studies have explored the influence of particles on interfacial dynamics, when the length scales of a liquid thread or film and the particle diameter are comparable. Because the particles deform the air-liquid interface, they strongly modify the liquid film geometry and instabilities \cite{furbank2004experimental,buchanan2007pattern,bonnoit2012accelerated,miskin2012droplet,lubbers2014dense,mathues2015capillary,kim2017formation,yu2017armoring,yu2018separation}. In the dip-coating configuration, only a few studies have investigated the fate of individual non-Brownian particles in the meniscus and their entrainment in the film \cite{Ouriemi:2013hn,gans2019dip}. An experimental study by Kao \textit{et al.} focused on the self-assembly of a monolayer of entrained particles in the liquid film \cite{kao2012spinodal} and bubbles \cite{kao2010pulling}. Recently, Colosqui \textit{et al.} considered 2D numerical simulations of a small number of dilute particles and showed the existence of two regimes of entrainment that result in the formation of clusters of particles, either within the meniscus or in the liquid film \cite{colosqui2013hydrodynamically}. They also suggested that no particles are entrained in the liquid film below a critical withdrawal velocity, corresponding to a coating film thinner than the particle diameter. However, no systematic validation of this prediction has been performed. In addition, the validity of the entrainment condition provided by Colosqui \textit{et al.} remains to be validated experimentally. Numerically, the cost of the 2D (and 3D) numerical simulations is too high to explore the validity of the entrainment condition. In addition, most applications involve 3D particles (spheres, aggregate, etc.) whose shape can influence the threshold for particle entrainement and the formation of clusters. There is therefore a need of systematic experiments to determine the conditions under which a particle is entrained in a liquid film during a dip coating process.

\medskip

In this paper, we show experimentally that capillary effects can prevent sufficiently large particles from being entrained in the liquid film during the dip coating process. We also demonstrate that particles up to six times larger than the liquid film can contaminate the withdrawn surface and we show the relevance of our findings for microorganisms. In section II, we first present our experimental approach: a glass plate is withdrawn from a liquid bath containing dispersed particles. In section III, we highlight the existence of three regimes for increasing withdrawal velocity: one where the coating film is free of particles, one in which the film contains individual isolated particles and one in which the entrained particles are assembled into clusters. We rationalize these observations in section IV and provide a discussion of the scaling law that describes the entrainment. Finally, in section V, we demonstrate that the capillary filtration mechanism is an original and effective way to prevent the contamination of a substrate dipped in a polluted liquid, containing particles or microorganisms. 

\section{Experimental methods}

To study the entrainment of non-Brownian particles in a coated film, we withdraw a glass plate from a bath of viscous dilute suspension at constant velocity $U$. The plate entrains a liquid film whose thickness in the absence of particles obeys the classical Landau-Levich law \cite{landau1942physicochim,maleki2011landau,rio2017withdrawing}.

The experiments are performed with density-matched polystryrene particles dispersed in silicone oil. We use high density silicone oil, AP 100, AR 200 or AP 1000 (from Sigma Aldrich). Because of possible small temperature variations during the experiments, we measure the evolution of the viscosity and the density of the silicone oils in the temperature range $T\in[16^{\rm o}{\rm C},\, 26^{\rm o}{\rm C}]$ with an Anton Paar MCR 501 rheometer. The dynamic viscosity for the silicone oil AP 100 is captured by the expression $\eta_{100}=0.1315-0.0051\,(T-20)$ (in Pa.s), for the AR 200: $\eta_{200}=0.2425-0.0077\,(T-20)$ and for the AP 1000: $\eta_{1000}=1.4227-0.0669\,(T-20)$, where $T$ is the temperature in degree Celsius. The density can be calculated with the following expressions, for the silicone oil AP 100: $\rho_{100}=1.0626-0.0008\,(T-20)$ (in ${\rm{g.cm^{-3}}}$), for the AR 200: $\eta_{200}=1.046-0.0008\,(T-20)$ and for the AP 1000: $\eta_{1000}=1.0871-0.0.0008\,(T-20)$, where $T$ is the temperature in degree Celsius. The silicone oils have an interfacial tension at $20^{\rm o}{\rm C}$ of $\gamma=21 \pm 1\,{\rm mN.m^{-1}}$. The polystyrene particles of radius $a=[20,\,40, \,70,\,125,\,250] \,{\rm \mu m}$ and average density of $\rho_p = 1055\, {\rm kg\,m^{-3}}$ (Dynoseeds TS - Microbeads) are dispersed in the liquid. Between two experiments the suspension was thoroughly mixed and over the timescale of an experiment, typically between a few seconds to a few tens of minutes, the suspension can be assumed to be neutrally buoyant. Indeed, the settling velocity of a particle in a quiescent fluid is due to a density mismatch between the fluid and the particle, and is given by \cite{lamb1993hydrodynamics}
\begin{equation}
v=\frac{2}{9}\,\frac{\rho_f\,\Delta \rho\,g\,a^2}{\eta},
\end{equation}
where $\Delta \rho=\rho_p-\rho_f$ is the density difference between the particle and the fluid, $\nu$ is the dynamic viscosity, $g$ is the gravitational acceleration, and $a$ is the radius of the particle. We compare the gravity-induced displacement of the particles with the distance traveled by the plate during an experiment. We found that this gravity-induced displacement is negligible for all suspensions and entrainment velocities. Therefore, the suspension can be considered as neutrally buoyant over the timescale of the experiment. 

\medskip

 The volume fraction of particles in suspension $\phi$ is defined as the ratio of the volume of particles to the total volume of fluid. We work with dilute suspensions, \textit{i.e.}, $\phi < 5\,\%$. We can thus neglect the collective effects associated with dense mixtures, including their non-Newtonian rheology. In particular, we assume that the presence of particles does not significantly modify the viscosity of the dilute suspension. This approximation holds for many situations, including samples with biological contaminants, typically at very low concentrations.

\medskip

\begin{figure}
\begin{center}
 \includegraphics[width=\textwidth]{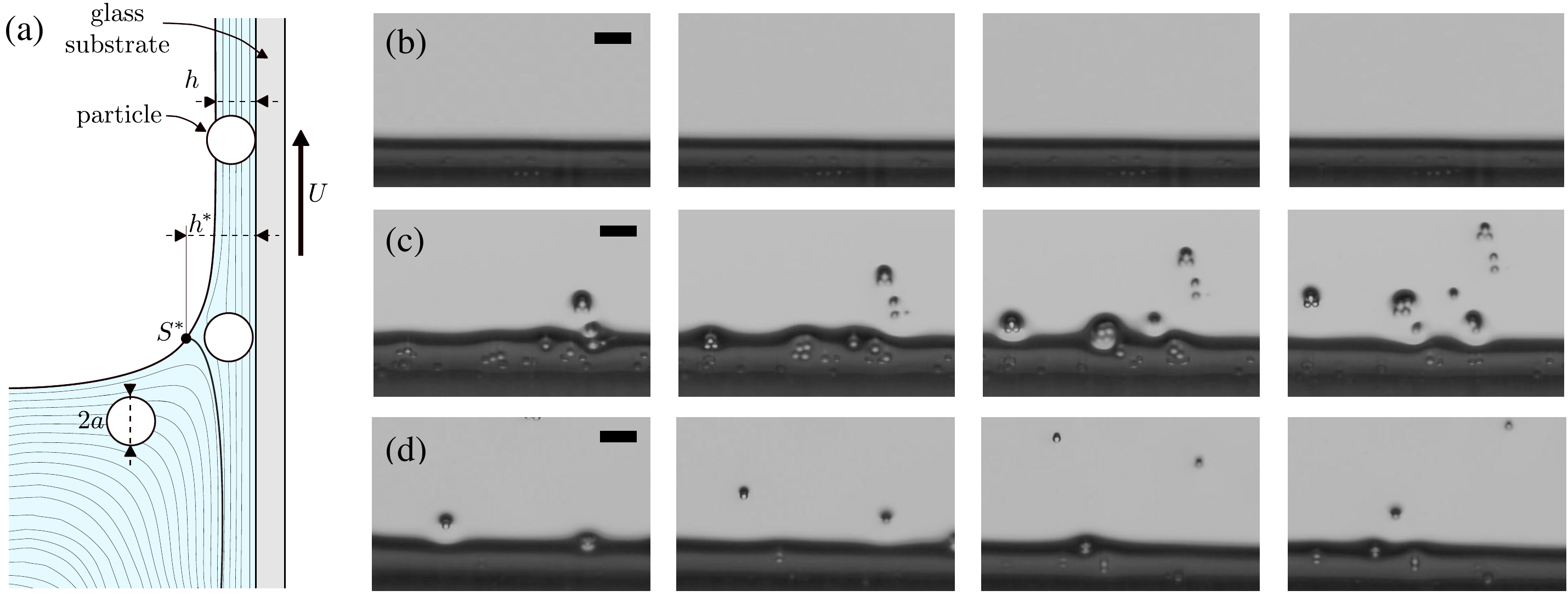}%
 \caption{(a) Schematic of the dip coating experimental setup. Experimental visualizations of the three entrainment regimes during the withdrawal of the substrate from a suspension (2\% of PS particles of radius $a = 125\,\mu{\rm m}$ in AP 100). (b) liquid only, (c) entrainment of clusters and (d) entrainment of individual particles. The time increases from left to right. Scale bars are $500\,\mu{\rm m}$\label{Figure1_Setup}}
 \end{center}
 \end{figure}

Once homogeneous, the suspension is transferred to a container (110 mm long $\times$ 40 mm wide $\times$ 150 mm high) whose vertical motion is actuated by a stepper motor (Zaber), with a speed up to $2.4 \,{\rm cm \, s^{-1}}$ with a $\pm 2\,\%$ precision. The solid substrate, a glass plate of thickness $l=2\,{\rm mm}$, width $w=100\,{\rm mm}$ and height $h=120\,\rm{mm}$ is initially dipped in the suspension and remains stationary during the experiment to prevent vibrations from disturbing the coated film. Prior to the experiments, the glass plates were thoroughly cleaned with Isopropanol (IPA) and dried with compressed air. A new container was used for each set of experiments. As the container moves downward, the plate emerges from the bath at a constant velocity $U$ and is coated with a liquid film over the 6.5-cm vertical course of the motor. A schematic of the experimental setup is shown in Fig. \ref{Figure1_Setup}(a). The dip coating dynamics, including the film thickness and composition are recorded with a digital camera (Nikon D7100) and a 200-mm macro lens focused above the meniscus. To detect the presence of the smallest particles ($a= 20 \,{\rm \mu m}$), we also use a long working distance microscope lens (10X EO M Plan Apo Long Working Distance Infinity Corrected from Edmund Optics).

 \medskip

The thickness of the liquid film without particles is determined using a gravimetry method. The volume of suspension removed from the container during a withdrawal can be obtained from the difference of mass measured by a scale $\delta m$ and the fluid density $\rho$, with $V=\delta m/\rho$. The main issue with this method is the presence of a lower edge effect \cite{Ouriemi:2013hn}. Therefore, we perform two withdrawal experiments with different dipping lengths $l_2>l_1$. Subtracting the masses obtained with the two dipping lengths allows defining and subtracting the volume of fluid that remains attached to the bottom of the plate after the withdrawal and we obtain the average fluid thickness $h$. In addition to the gravimetry methods used to measure the thickness of the liquid film in the pure silicone oil, we perform measurements of the film thickness using a light absorption method, similarly to Vernay \textit{et al} \cite{vernay2015free}. When the light passes through the coated liquid film, its absorption is proportional to the thickness of the film following the Beer-Lambert law of absorption. We first perform a careful calibration through samples of controlled thickness and determine that for the Sudan Red and the silicone oil used (AP 100), the intensity response of the Sudan Red matches the Beer-Lambert law of absorption $h={\rm log}\left(I/I_0\right)/(\varepsilon\,c)$. 


\section{Results}

\begin{figure}
 \subfigure[]{\includegraphics[width=0.45\textwidth]{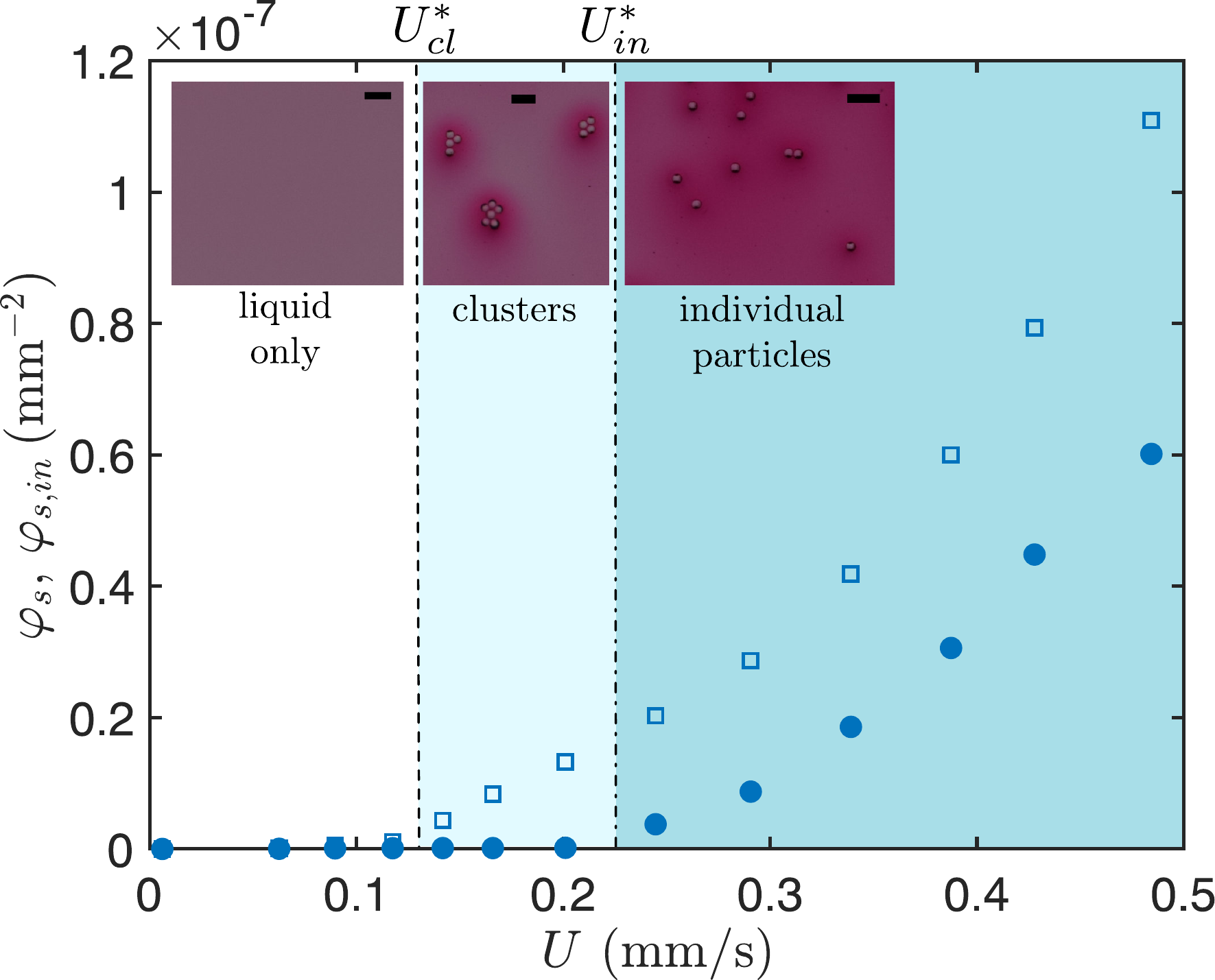}} \quad
  \subfigure[]{\includegraphics[width=0.45\textwidth]{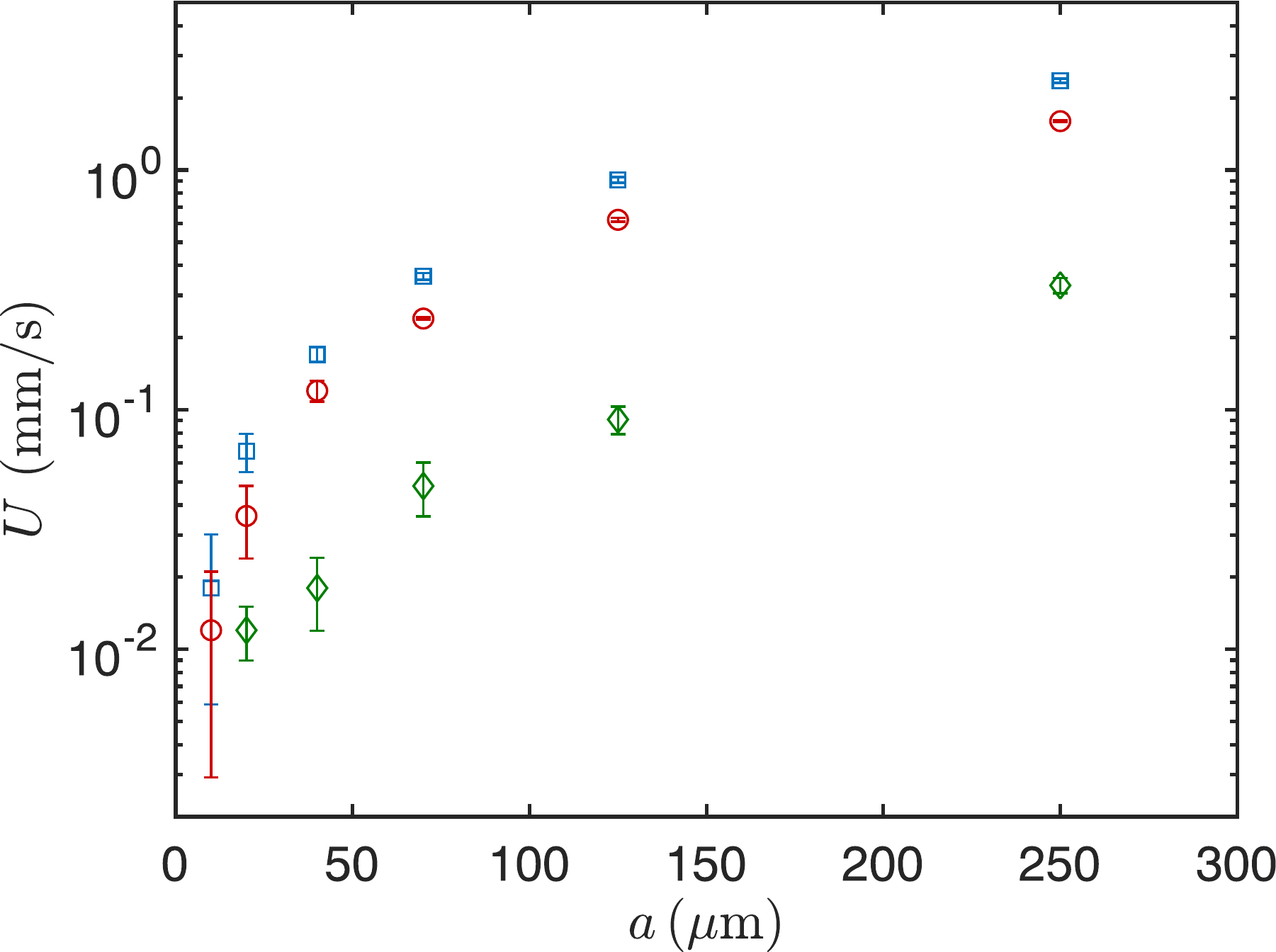}}
 \caption{(a) Evolution of the surface density of particles entrained $\varphi_s$ (hollow blue rectangles) and individual particles entrained $\varphi_{s,in}$ (blue filled circles) when increasing the withdrawal velocity $U$ for a suspension of polystyrene particles dispersed in silicone oil AP 100 ($\phi=5\,\%$ and $a=70\,\mu{\rm m}$). The vertical dashed line and dash-dotted line show the threshold velocities for clusters entrainment $U^*_{cl}$ and individual particle entrainement $U^*_{in}$, respectively. The insets illustrates the composition of the coated film in the different regimes. Scale bars are $500\,\mu{\rm m}$. (b) Threshold velocity $U_{in}^*$ for the entrainment of individual particles when varying the radius of the particles $a$ and the liquid phase: silicone oil AP 100 (blue squares), AR 200 (red circles), AP 1000 (green diamonds). The experiments are performed only for dilute suspensions, $\phi<0.5\%$.\label{Figure2_Setup}}
 \end{figure}
 
\subsection{Phenomenology}

We begin by examining the entrainment of particles in the liquid film for different withdrawal velocities. We observe three entrainment regimes when increasing the withdrawal velocity $U$ [Fig. \ref{Figure1_Setup}(b)]. At low withdrawal velocity, no particles are entrained in the film. The particles are trapped in the meniscus, which effectively acts as a capillary filter. At large withdrawal velocity, isolated particles are entrained in the liquid film and contaminate the substrate. The thickness of the liquid film increases locally around the particles, which modifies the properties and stability of the liquid coating. At intermediate velocity, clusters or aggregates of particles are entrained in the liquid film. The assembly of the particles, \textit{i.e.}, the formation of the clusters takes place in the meniscus. Indeed, the meniscus traps individual particles. When the local density of particles is large enough, clusters form, cross the meniscus and get entrained in the film. The withdrawal velocity at which clusters are first observed depends on the probability of the particles to assemble in the meniscus and thus on the volume fraction of the suspension and the withdrawal length. 
Based on these qualitative observation, the capillary filtration, which is responsible for trapping particles in the meniscus, occurs at low withdrawal velocity where it prevents contamination of the thin liquid film and contributes to the formation of clusters at intermediate withdrawal velocity. This mechanism is not relevant at high withdrawal velocity. 

\subsection{Entrainement threshold}

To better understand the physical mechanisms involved in capillary filtration and determine the conditions under which particles are not entrained in the liquid film, we carry out systematic experiments varying the withdrawal speed and the properties of the suspension. Here, we consider a dilute suspension ($\phi < 5\%$). The experiment consists in increasing the withdrawal velocity while keeping other parameters fixed and recording the number of particles entrained per unit surface area of film [Fig. \ref{Figure2_Setup}(a)]. We record the total number of particles per unit area $\varphi_s$ and the total number of individual isolated particles $\varphi_{s,in}$, \textit{i.e.}, particles that are not in a cluster. At low withdrawal velocity, we observe that no particles are entrained in the liquid film whose thickness remains well predicted by the Landau-Levich law. Increasing the withdrawal velocity leads to a first transition at $U_{cl}^*$ above which only clusters are entrained in the liquid film, and no individual particles are observed. Finally, increasing the withdrawal velocity further, beyond $U_{in}^*$, leads to the entrainment of individual particles. We first focus on the situation when individual particles are entrained in the liquid film, and we shall discuss later the entrainment of clusters. To focus on the individual particle regime, we consider very diluted suspensions ($\phi<0.5\%$). Indeed, at low particle concentration, the probability to assemble clusters in the meniscus is low. The efficiency of the filtration process drops only slightly when clusters start being entrained, but then increase suddenly as isolated particles are entrained. 

 \begin{figure}
 \begin{center}
  \subfigure[]{\includegraphics[width=0.45\textwidth]{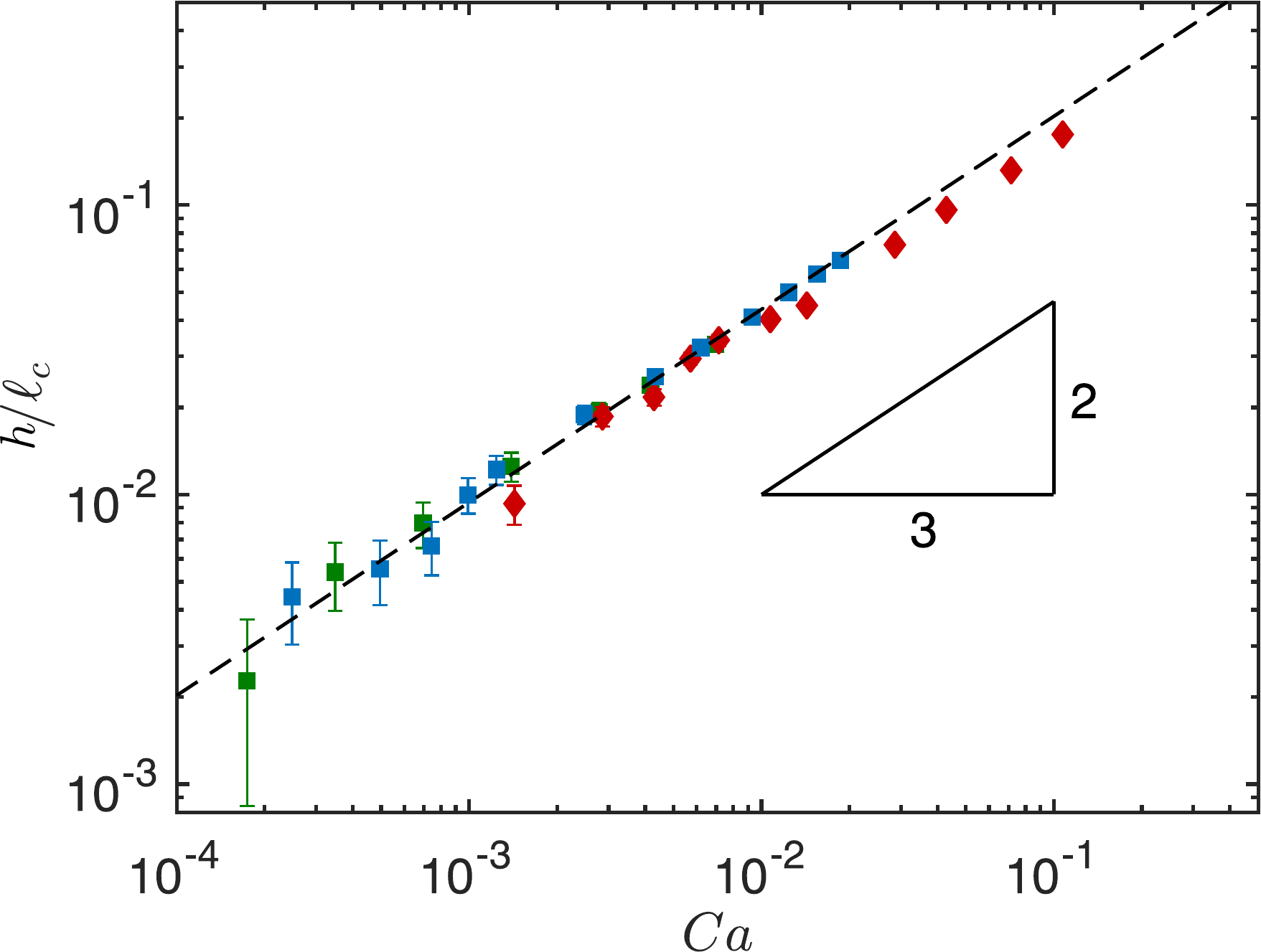}} \quad
   \subfigure[]{\includegraphics[width=0.45\textwidth]{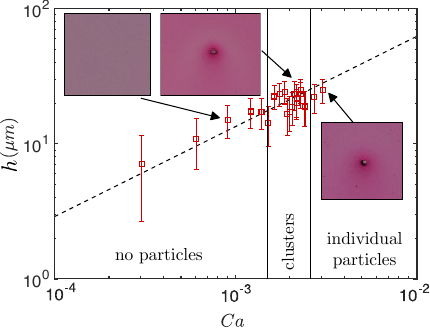}} 
 \caption{(a) Rescaled thickness of the liquid film $h/\ell_c$ without particles for varying capillary number $Ca$ and silicone oil AP 100 (green sybmols), AR 200 (blue symbols), AP 1000 (green symbols). The dotted line is the theoretical prediction $h/\ell_c=0.94\,Ca^{2/3}$. (b) Thickness of the liquid film when varying the capillary number and withdrawing the glass plate from a suspension of polystyrene particles in silicone oil (AP 100) with $\phi=0.28\,\%$ and $a=70\,\mu{\rm m}$ particles. The vertical lines indicate the three different regimes and the insets show typical film compositions on the plate. \label{Fig_2}}
  \end{center}
 \end{figure}

Using particles of different radii, we measure the threshold velocity $U_{in}^*$ at which individual particles are entrained in the liquid film, and we show that the larger the particle is, the greater the threshold velocity is [Fig. \ref{Figure2_Setup}(b)]. The capillary filtration is more efficient for large particles as they are trapped in the meniscus for a larger range of withdrawal velocity. The properties of the fluid also determine the efficiency of the capillary filtration. Using different oils, we vary the viscosity of the fluid. Increasing the viscosity of the fluid decreases the threshold velocity $U_{in}$ for all particle sizes [Fig. \ref{Figure2_Setup}(b)]. The less viscous the liquid is, the more efficient the capillary filtration.

For dilute suspensions, the entrainment of particles depends on the thickness of the coated film. The thickness of liquid deposited on a plate withdrawn from a bath of Newtonian fluid scales with a power-law of the capillary number $Ca= \eta \, U / \gamma $ in the limit of small capillary numbers, typically $Ca<10^{-2}$ \cite{maleki2011landau}. The thickness of the liquid layer $h$ is uniform along the plate and follows the Landau-Levich-Deryaguin law (LLD law): $ h = 0.94 \, \ell_c \, Ca^{2/3} $, where $ \ell_c = \sqrt{\gamma/(\rho\,g)}$ is the capillary length of the fluid. We performed dip coating experiments with silicone oils, without particles, and measured the film thickness by classical gravimetry methods \cite{maleki2011landau,krechetnikov2006surfactant,Ouriemi:2013hn}. Our measurements are reported in Fig. \ref{Fig_2}(a) and show an excellent agreement with the classical LLD law in the range $Ca\in [10^{-4},2\,\times 10^{-2}]$ for all silicone oils. For dilute suspensions, we will, therefore, use the LLD model to describe the thin film and the forces acting on the particles.

 Experimentally, we observe that the presence of a small amount of particles does not modify significantly the thickness of the coating film sufficiently far from the particles [Fig. \ref{Fig_2}(b)]. Indeed, experiments performed with a diluted suspension ($\phi=0.28\,\%$ of $2\,a=140\,\mu{\rm m}$ particles in silicone oil AP 100) show that we recover the classic Landau-Levich law in all regimes, far from the particles, as reported in Fig. \ref{Fig_2}(b).


 \section{Discussion}

 \subsection{Threshold of particles entrainement}

\begin{figure}
\includegraphics[width=0.75\textwidth]{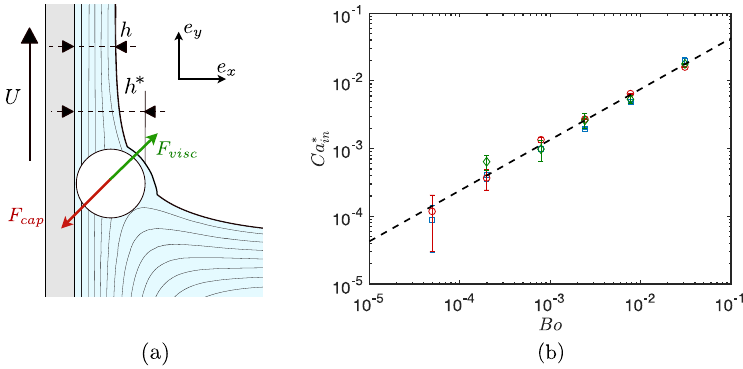}
 \caption{(a) Schematic of the forces acting on a particle in the liquid meniscus. (b) Threshold capillary number $Ca_{in}^*$ for individual particle entrainement in the liquid film. The symbols are the experimental results for different silicone oils (blue squares: AP 100, red circles: AR 200, green diamonds: AP 1000) and the dotted line is the scaling $Ca_{in}^* =0.24\,{\rm Bo}^{3/4}$. \label{Fig_22}}
 \end{figure}

 The condition of filtration depends on the forces acting on a particle in the meniscus, the capillary and viscous forces, as represented in Fig. \ref{Fig_22}(a). The viscous force is responsible for the entrainment of the particles in the liquid film whereas the capillary force which opposes the deformation of the air/liquid interface prevents the particles from entering the liquid film, hence leading to capillary filtration. Colosqui \textit{et al.} suggested that a 2D particle is entrained when the thickness of the liquid film at the stagnation point, $h^*$, is larger than the particle size.

During the dip coating of a plate, the thickness at the stagnation point is given by $h^*/\ell_c=\left[3\,(h/\ell_c)-(h/\ell_c)^3/Ca\right]$ \cite{landau1942physicochim,krechetnikov2010application}. For a particle to be entrained in the liquid film, the viscous drag force on the particle needs to be larger than the resistive capillary force. An upper bound for the entrainment threshold is given by the condition $h^*>2\,a$. Thus, the scaling for particle entrainment, with $h^* \simeq 3\,h= 3\,e=2.82\,\ell_c\,Ca^{2/3}$, becomes
\begin{equation}\label{eq_seuil}
Ca^* = 0.59\,Bo^{3/4},
\end{equation}
where $Bo=\left(a/\ell_c\right)^2$ is the Bond number. This condition is derived based on geometrical criteria and is the same as the one suggested by Colosqui \textit{et al.} \cite{colosqui2013hydrodynamically}. The geometrical condition is equivalent to assuming that the particle is entrained in the liquid film when the capillary force vanishes. For capillary numbers greater than $Ca^* $, the capillary force is not sufficient to filter out individual particles, \textit{i.e.}, to prevent them from being entrained in the liquid film.

Our experiments reveal that the scaling law captures the trend, as a $3/4$ power law of the Bond number, but the prefactor is not correct. In Fig. \ref{Fig_22}(b), we report the threshold value of the capillary number as a function of the Bond number for the experiments originally presented in Fig. \ref{Figure2_Setup}(b). All the data points collapse on the line corresponding to the scaling law $ Ca_{in}^* \propto Bo^{3/4} $, for $ Bo $ varying over three orders of magnitude. This power law is consistent with the model. The prefactor measured experimentally is equal to 0.24, and differs from the value obtained in the model. This difference likely comes from the condition that the capillary force has to vanish is too restrictive. This indicates that the force balance between the capillary effects and viscous force needs to be refined to
predict accurately the condition of entrainment. As a quantitative model for the capillary force in the regime of large deformation of the meniscus is still missing, the experiments presented in this paper are the only method to quantitatively determine the entrainment threshold.

 \subsection{Entrainement of clusters}

\begin{figure}
 \includegraphics[width=1\textwidth]{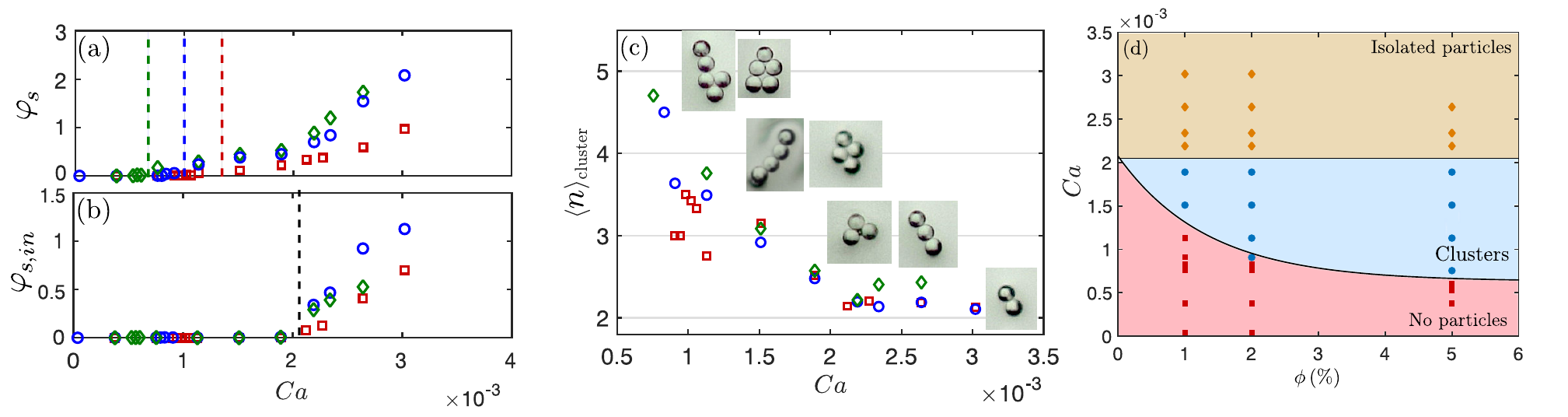}%
 \caption{Evolution of the surface density of (a) the total particles entrained $\varphi_s$ and (b) only the individual particles entrained $\varphi_{s,in}$ when varying the capillary number $Ca$ and for different volume fractions $\phi=1\%$ (green diamonds), $\phi=2\%$ (blue circles) and $\phi=5\%$ (red squares). In (a) the colored dashed lines show the transition between the entrainment of liquid alone and the entrainment of particles in clusters. In (b) the vertical line shows the transition between the cluster regime and the individual particle regimes. (c) Mean size of the clusters entrained in the liquid film. The insets show different morphologies of clusters. (d) Diagram $(\phi,\,Ca)$ reporting the existence of the three regimes: liquid alone, clusters and individual particles, obtained for $\Delta= = 33\,{\rm  mm}$ and various volume fraction $\phi$. The black lines are guides for the eye.\label{Fig_3}}
 \end{figure}

In the previous section, we reported on the entrainment of individual particles in very dilute suspensions ($ \phi <0.5 \% $). Depending on the particle volume fraction, and the length over which the plate is withdrawn from the suspension, $\Delta$, we also observe an intermediate regime, in which clusters of particles are entrained. The clusters assemble in the meniscus and cross into the film for $Ca<0.24  \, Bo^{3/4}$, before the entrainment of individual particles. The formation of clusters in the meniscus effectively decreases the efficiency of the filtration process. To determine the conditions of cluster formation and entrainment in the liquid film, we conduct withdrawal experiments with suspensions of 140 $\mu$m particles in the AP 100 silicone oil. For different volume fractions of particles in suspension $ 0.3 \% <\phi <5 \% $, we measure the surface fraction of particles entrained in the liquid film $\varphi_s$ over a constant withdrawal distance of the plate $ \Delta=33 $ mm. We also measure the surface fraction of individual particles $ \varphi_{s,in}$ [Fig. \ref{Fig_3}(a)-(b)]. We define two entrainment thresholds, one for the clusters and one for individual particles using $ \varphi_{s, in}$ and $ \varphi_s = \varphi_{s, cl} + \varphi_{s, in} $, where $ \varphi_{s, cl}$ is the surface fraction of particles in clusters.

The results indicate that the threshold capillary number at which individual particles are entrained on the plate does not depend on the volume fraction $\phi$ of the suspension. For the suspensions used here, $Ca_{isol}^* \simeq 2.2\times10^{-3}$ for all particle densities [Fig. \ref{Fig_3}(b)]. This result is consistent with the model presented in the previous paragraph: $Ca^*$ only depends on $Bo$, which is independent of $\phi$.

In contrast, the threshold capillary number above which clusters are entrained in the liquid film decreases as the volume fraction of the suspension $\phi$ increases [Fig. \ref{Fig_3}(a)]. In this regime, the individual particles that are not able to enter the liquid film are joined by other particles being advected along streamlines toward the meniscus, thus forming clusters driven by capillary forces in the vicinity of the stagnation point \cite{kralchevsky2000capillary,vella2005cheerios,stebe2009oriented,cavallaro2011curvature}. The formation of clusters both distorts the interface locally and increases the viscous drag on the cluster. Above a certain velocity, when the probability of forming a cluster of a given size is sufficient, clusters are able to enter the film. The efficiency of the capillary filtration is therefore dependent upon the particle concentration. Indeed, for a given volume of suspension entrained, the number of particles trapped in the meniscus increases with the volume fraction of the suspension $\phi$ and the withdrawal length $\Delta$, making this process time dependent. A greater number of particles can then assemble into clusters in the vicinity of the stagnation point during the withdrawal of the plate from the bath of suspension. It is important to note that the number of particles that can assemble in the meniscus not only depends on the volume fraction $\phi$ but also on the withdrawal length of the plate $\Delta $. We report experiments performed with a constant withdrawal length $ \Delta = 33 \, {\rm mm} $, but increasing the value of $ \Delta $ leads to more particles collected in the vicinity of the stagnation point and decreases the threshold capillary number for clusters entrainment. The trapping of particles in the meniscus not only increases the probability for particles to be in a cluster but it can also determine the size and shape of those clusters. Increasing the length also allows for more time for entrained particles to aggregate within the film, due to capillary forces.

We measure the average size of the clusters entrained on the substrate as a function of the capillary number $Ca$ for different values of the volume fraction of the suspension $\phi$. The results are reported in Fig. \ref{Fig_3}(c) and show that, on average, the larger the clusters are, the lower the capillary number at which they are entrained is. Large clusters are entrained far below the threshold value obtained for individual particles $Ca_{in}^*$. Indeed, clusters composed of 4 to 5 particles are entrained at capillary numbers that are equal to about half the threshold value to entrain isolated particles. The threshold capillary numbers obtained for clusters of different sizes are independent of the concentration of the suspension. Only the probability to assemble a cluster of a fiven size in the meniscus varies with $\phi$. The quantitative prediction of the entrainment of a cluster, depending on its size and shape, requires an analytical expression for the capillary force. Qualitatively, the clusters are composed of a single layer of particles and the presence of a larger number of particles in a cluster increases its surface area and the viscous lubrication force whereas the capillary force depends on the width of the cluster. This allows the cluster to overcome the capillary filtering mechanism induced by the meniscus at a smaller film thickness than individual particles. We also note that the amount of particles entrained on the plate for  $Ca<Ca_{in}^*$ is much smaller than when the threshold $ Ca \simeq 0.24 \, Bo^{3/4} $ is reached, as the majority of particles remain isolated in the meniscus [Fig. \ref{Fig_3}(a)].

Finally, we summarize our findings in a phase diagram of the three regimes as a function of the capillary number $Ca$ and the volume fraction $\phi$ in Fig. \ref{Fig_3}(d). The threshold for individual particle entrainment, which is the main entrainment mechanism for very dilute suspensions, is independent of the volume fraction $\phi$. However, for more concentrated suspensions, capillary filtration is limited by the formation of clusters which depends on $\phi$ and on the withdrawal length $\Delta$. Therefore, a substrate withdrawn from a liquid bath containing particles is more likely to be contaminated at lower withdrawal velocities if the volume fraction of the suspension, or the amount of contaminants, is substantial.

 \subsection{Application to biological contamination}

  In first approximation, the capillary filtration mechanism described here can be extended to biological organisms. The threshold capillary number or velocity is useful information to prevent the contamination of an object removed from a liquid bath containing dilute microorganisms. This filtration mechanism can be valuable in biological applications in which a probe is dipped in water containing bacteria or microalgae. 

\begin{figure}[h]
 \includegraphics[width=0.5\textwidth]{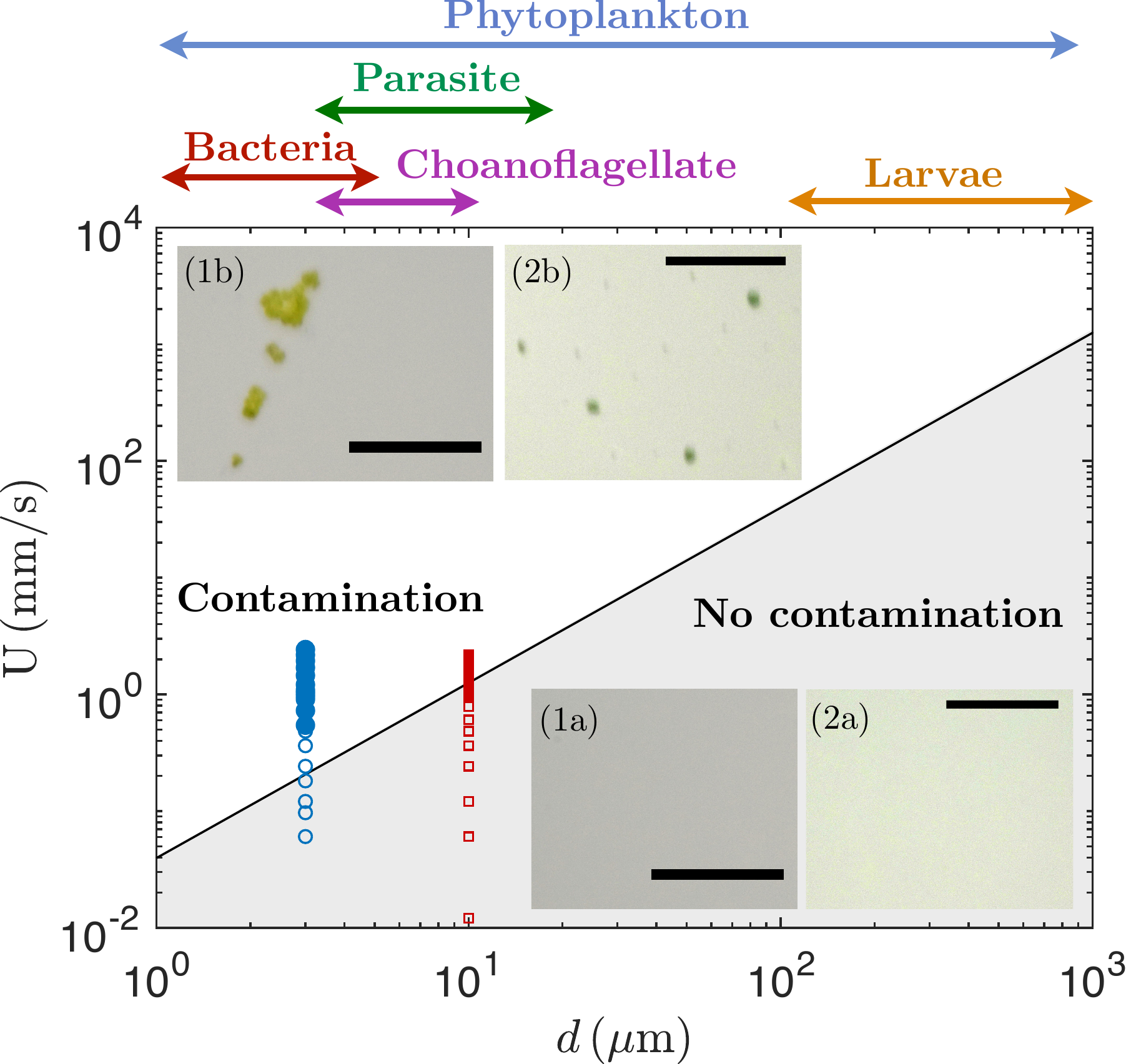}%
 \caption{Velocity threshold leading to the contamination of a substrate withdrawn from contaminated water (white region). In the grey region, the substrate is expected to be withdrawn from the polluted water without contamination thanks to the capillary filtering mechanism. The black corresponds to Eq. \ref{eq_dimensional}. The arrow indicates the range of size of different biological organisms \cite{dervaux2017light,guasto2012fluid,mino2017finding,rusconi2014bacterial,stocker2012marine,vesey1994application}. The filled symbols and the hollow symbols indicates contaminated and non-contaminated substrates, respectively. The blue circles are experiments with \textit{Synechocystis sp.} PCC6803 and the red symbols are experiments with \textit{Chlamydomonas reinhardtii}. The insets photographs are obtained from withdrawal experiments from suspensions of \textit{Chlamydomonas reinhardtii} ($2\,a \sim 10\,\mu\rm{m})$ [(1a)-(1b)] and \textit{Synechocystis sp.} PCC6803 ($2\,a \sim 3\,\mu\rm{m})$ [(2a)-(2b)]. At low withdrawal velocity, the micro-organisms are not entrained: for (1a), $U=0.36\,{\rm mm.s^{-1}}$ and for (2a) $U=0.48\,{\rm mm.s^{-1}}$. At larger withdrawal velocity, the micro-organisms contaminate the substrate: for (1b), $U=2.18\,{\rm mm.s^{-1}}$ and for (2b) $U=1.03\,{\rm mm.s^{-1}}$. The scale bars are 200$\mu$m.\label{Fig_bio}}
 \end{figure}

Biological species are usually very diluted in water, so we only consider the threshold obtained for the entrainment of individual particles. Assuming the shape of biological organisms is spherical, we can use the relation obtained previously, $ Ca_{in}^* \simeq 0.24 \, Bo^{3/4}$ in terms of dimensional quantities:
\begin{equation}\label{eq_dimensional}
U^*= 0.24\,\frac{\rho^{3/4}\,g^{3/4}\,\gamma^{1/4}}{\eta}\,a^{3/2}.
\end{equation}
Water has a surface tension $\gamma = 72 \, {\rm mN/m}$, a density $\rho=1000\,{\rm kg/m^3}$ and a dynamic viscosity $\eta = 10^{-3} \, {\rm Pa.s}$.

In Fig. \ref{Fig_bio} we plot the threshold velocities at which biological contaminants are expected to be entrained in the liquid film on the flat plate based on physical values found in the literature 
\cite{dervaux2017light,guasto2012fluid,mino2017finding,rusconi2014bacterial,stocker2012marine,vesey1994application,jana2015somersault,durham2012thin}. For microorganisms such as bacteria and parasites, the substrate needs to be pulled out of the polluted water at velocities smaller $10^{-2}-10^{-1}\,({\rm mm.s^{-1}})$ to prevent contamination. For larger microorganisms such as phytoplankton and larvae, the threshold speed increases, of the order of a few tens of centimeters per second. These threshold values, under which capillary filtration prevents contamination, can provide guidelines to operate probes and mixers and for rinsing processes in biological and industrial applications. 

To obtain a better estimate of the threshold velocity at which biological particles will contaminate a substrate, the shape, the deformability and the motility of the particles would need to be taken into account. However, the values derived for hard particles already give a good estimate of the threshold velocities for microorganisms. Indeed, we performed experiments with dilute suspensions of two different micro-organisms in water at usual volume fraction $\phi<0.1\%$: \textit{Chlamydomonas reinhardtii} and \textit{Synechocystis sp.} PCC6803. An illustration of the results is reported in the inset of Fig. \ref{Fig_bio} and shows that below the expected threshold, the substrate is withdrawn from the polluted water with no microorganism. The thresholds obtained with these two micro-organisms are in fair agreement with the theoretical prediction. The small discrepancies may be due to the shape, size distribution, and mechanical properties of the microorganisms. At velocity smaller than the expected threshold, the substrate is not contaminated by microorganisms.

\section{Conclusion}

During the withdrawal of a flat substrate from a liquid bath containing particles or biological micro-organisms, the meniscus can effectively act as a capillary filter that could be sometimes more effective than a passive filter as used in microfluidic devices that clog over time, reducing their efficiency \cite{Lenshof2010,sauret2014clogging,dressaire2017clogging,sauret2018growth}. Below a given film thickness on the plate, the meniscus generates a strong capillary force at the stagnation point and prevents the particles from being entrained in the liquid film and contaminating the substrate. Assuming that the thickness at the stagnation point is smaller than a fraction of the particles size, {\textit{i.e.}} that the capillary number is small enough, the particles are trapped in the liquid film for Capillary number smaller than a threshold value $Ca^*=0.24\,Bo^{3/4}$. For larger volume fraction, typically larger than a few percents, some trapped particles assemble into clusters that can be entrained in the liquid film. The capillary filtering mechanism described here should ,apply to a wide range of quasi-spherical particles, including biological micro-organisms. Besides, the threshold dependence on the particle size could allow particle sorting by size using the same filtration mechanism.

 \section*{Acknowledgements}

We thank H. A. Stone, C. Colosqui, and G. M. Homsy for important discussions and P. Brunet for providing the \textit{Chlamydomonas reinhardtii} and the \textit{Synechocystis sp.} PCC6803. This work was supported by the French ANR (project ProLiFic ANR-16-CE30-0009) and partial support from a CNRS PICS grant n$^{\rm o}$07242.

\bibliography{Biblio_DipCoating_Diluted}

\end{document}